\documentclass[preprint,aps,draft]{revtex4}

\usepackage{bm}

\begin{document}

\preprint{IZTECH-P-2006-03}

\title{A symmetry for vanishing cosmological constant: Another 
realization}

\author{Recai Erdem}
\email{recaierdem@iyte.edu.tr}
\affiliation{Department of Physics,
{\.{I}}zmir Institute of Technology \\ 
G{\"{u}}lbah{\c{c}}e K{\"{o}}y{\"{u}}, Urla, {\.{I}}zmir 35430, 
Turkey} 

\date{\today}

\begin{abstract} A more conventional realization 
of a symmetry which had been proposed towards the solution of cosmological 
constant 
problem is considered. In this study the multiplication of the coordinates 
by the imaginary number $i$ in the literature is replaced by the 
multiplication of the metric tensor by minus one. This realization 
of the symmetry as well forbids a bulk cosmological constant and selects 
out $2(2n+1)$ dimensional spaces. On contrary to its previous 
realization the symmetry, without any need 
for its extension, also forbids a 
possible 
cosmological constant term which may arise from the extra dimensional 
curvature scalar provided that the space is taken as the union of two 
$2(2n+1)$ dimensional spaces where the usual 4-dimensional space lies at the 
intersection of these spaces. It is shown that this 
symmetry may be realized through spacetime reflections that change the 
sign of the volume element. A possible 
relation of 
this symmetry to the E-parity 
symmetry of Linde is also pointed out.
\end{abstract}

\maketitle

Recently a symmetry \cite{Erdem,Nobbenhuis,tHooft} which may 
give insight to the origin of 
the extremely small value \cite{PDG} of the cosmological 
constant compared to its theoretical value \cite{Wein} was proposed. As in 
the usual 
symmetry arguments the symmetry forces the cosmological constant vanish 
and the small value of the cosmological constant is attributed to the 
breaking of the symmetry by a small amount. In \cite{Erdem} the symmetry 
is realized by imposing the 
invariance 
of action functional under a transformation where all coordinates are 
multiplied by the imaginary number $i$. It was found that this symmetry 
select out the dimensions $D$ obeying $D=2(2n+1)$ $n=0,1,...$, that is, 
$D=2,6,10,....$ and it gives some constraints on the form of the possible 
Lagrangian terms as well. Moreover that symmetry has more chance to 
survive in quantum field theory when compared to the usual scaling 
symmetry because the n-point functions are 
invariant under this symmetry. In this paper we study a symmetry 
transformation where the 
coordinates remain the same while the metric tensor is multiplied by minus 
one. We show that this symmetry is equivalent to the one given in 
\cite{Erdem}.  Although its results are mainly the same as 
\cite{Erdem} it is more conventional in its form, in the sense that the 
space-time coordinates remain real. On contrary to \cite{Erdem} we use 
the same symmetry to forbid 4-dimensional cosmological constant as well as 
to forbid a bulk cosmological constant.  
Moreover we show that the multiplication 
of the metric tensor by minus one may be related to a parity-like 
symmetry in the extra dimensions. 
We also discuss the relation of 
this symmetry  to the anti-podal symmetry of Linde 
\cite{Linde,Kaplan,Moffat}, whose relation to the previous realization of 
the present symmetry is discussed also in \cite{tHooft} for the 
4-dimensional case.

The symmetry principle given in \cite{Erdem} may be summarized as follows: 
The transformation
\begin{eqnarray}
&& x_A \rightarrow i\, x_A
   \label{a1} \\
\mbox{implies}
~~~~~~~~&&R \rightarrow -R~~~~,~~~~~~ 
\sqrt{g} \,d^Dx \rightarrow
(i)^D\,\sqrt{g} \,d^Dx 
\label{a2} \\
&&ds^2\,=\,g_{AB}dx^A\,dx^B 
~\rightarrow 
~~-\,ds^2 \label{a3}
\end{eqnarray}
where $A,B=0,1,2,....,D-1$ and $D=1,2,..$ is the dimension of the 
spacetime. The requirement of the invariance of the gravitational action 
functional 
\begin{equation}
S_R = \frac{1}{16\pi\,G}\int 
\sqrt{g} \,R,d^Dx \label{a3a}
\end{equation}
under (\ref{a1}) selects out the dimensions
\begin{equation}
D=2(2n+1)~~~,~~~~n=0,1,2,3,....
\label{a4}
\end{equation}
and forbids a bulk cosmological constant $\Lambda$ in the action
\begin{equation}
S_C = \frac{1}{16\pi\,G}\int \sqrt{g} \,\Lambda \,d^Dx \label{a5}
\end{equation}
Extension of this symmetry to the full action requires that the 
Lagrangian should transform in the same way as the curvature scalar, that is, 
\begin{equation}
{\cal L} \,\rightarrow\,
-{\cal L} ~~~~~~\mbox{as}~~~~~~
x_A \rightarrow i\, x_A~~~~,~~~~\mbox{for}~~~~ D=2(2n+1)~,~~~~~ 
A=0,1,....D-1
\label{a6}
\end{equation}
(i.e. for the dimensions given by Eq.(\ref{a4}) ). 
The kinetic terms of the scalar and vector fields automatically satisfy 
Eq.(\ref{a6}) while the potential terms ( e.g. $\phi^4$ term) are, 
in general, allowed in the lower dimensional sub-branes. The fermionic 
part 
of the Lagrangian does not satisfy (\ref{a6}) in general so fermionic 
fields may live only on a lower dimensional subspace (brane). For example 
the free fermion Lagrangian is allowed on a $4m+1$ dimensional subspace of 
the $2(2n+1)$ dimensional space, where $m\,\leq\,n$ $n,m=0,1,2, ...$. 
Although the transformation rules for the fields are similar to the ones 
for scale transformations ( where the scale parameter is 
replaced by the imaginary number $i$) this symmetry has a better chance 
of surviving after quantization because the two point functions ( e.g. 
$<0|T\phi(x)\phi(y)|0>$ for scalars and $<0|T\psi(x)\bar{\psi}(y)|0>$ for 
fermions), which are 
the basic building blocks for connected Feynman diagrams, are invariant 
under this symmetry transformation.

Now I introduce a symmetry transformation which is essentially 
equivalent to (\ref{a1}) while formulated in a more conventional form, 
that is,
\begin{equation}
g_{AB}\rightarrow -\,g_{AB} \label{a7}
\end{equation}
Eq.(\ref{a7}) induces 
\begin{eqnarray}
&&R \rightarrow -R~~~,~~~~~~ 
\sqrt{g} \,d^Dx \rightarrow
(\pm\,i)^D\,\sqrt{g} \,d^Dx \label{a8} \\
&&ds^2\,=\,g_{AB}dx^A\,dx^B 
~\rightarrow 
~~-\,ds^2 \label{a9}
\end{eqnarray}
The requirement of the invariance of the gravitational action 
(\ref{a3a}) under the transformation (\ref{a7}) 
selects out the dimensions given by
\begin{equation}
D=2(2n+1)~~~,~~~~n=0,1,2,3,....
\label{a10}
\end{equation}
as in Eq.(\ref{a4}), and for 
$D=2(2n+1)$, $n=0,1,2,3,...$ Eqs.(\ref{a8},\ref{a9}) become identical with 
Eqs.(\ref{a2},\ref{a3})
\cite{Duff}
. Moreover one notices that the requirement of the 
invariance of the action functional under (\ref{a7}) forbids a bulk 
cosmological constant term (given by (\ref{a5})) in $2(2n+1)$ dimensions. 
In other words the requirements of the invariance of the action functional 
under (\ref{a7}) and non-vanishing of its gravitational piece (\ref{a4}) 
implies $D=2(2n+1)$ and the vanishing of the bulk cosmological constant 
as in \cite{Erdem}. Although the implications of this symmetry 
for Lagrangian are similar to those of \cite{Erdem} there are some differences. We find 
it more suitable to consider this point after we consider the realization of 
this symmetry through reflections in extra dimensions in the 
paragraph after the next paragraph.

We have shown that the invariance of the gravitational action under 
Eq.(\ref{a7}) requires the vanishing of the bulk cosmological 
constant. The next step is to show that Eq.(\ref{a7}) results in  
the vanishing of the possible contributions due to extra dimensional 
curvature scalar as well so that the 
4-dimensional cosmological constant vanishes altogether. On 
contrary to 
\cite{Erdem} we use the same symmetry ( ( i.e. (\ref{a7}) ) that we have 
used to forbid the bulk cosmological constant) to forbid a possible 
4-dimensional cosmological constant induced by extra dimensional 
curvature scalar as well. To this end we take the 4-dimensional space-time 
be the intersection of two $2(2n+1)$ dimensional spaces; one with 
the dimension $2(2n+1)$ (e.g. 6) and the other with the dimension 
$2(2m+1)$ (e.g. 6) so that 
the total dimension of the space being $2(2m+1)+2(2n+1)-4=4(n+m)$ (e.g. 
8). 
Then Eq.(\ref{a7}) takes the following form  
\begin{eqnarray}
&&g_{AB}\rightarrow -\,g_{AB} ~~~~~,~~~~A,B= 
0,1,2,3,4^\prime.....D^\prime-1~~,~~
~D^\prime=2(2n+1) \label{a10a} \\
&&g_{CD}\rightarrow -\,g_{CD} ~~~~~,~~~~C,D= 
0,1,2,3,4^{\prime\prime},.....D^{\prime\prime}-1~~,~~
~D^{\prime\prime}=2(2m+1)
\label{a10b}
\end{eqnarray}
which transforms the metric and the curvature scalar as
\begin{eqnarray}
&&ds^2\,=\,g_{MN}dx^M\,dx^N =g_{\mu\nu}dx^\mu\,dx^\nu+g_{ab}dx^a\,dx^b 
\rightarrow\,
g_{\mu\nu}dx^\mu\,dx^\nu-g_{ab}dx^a\,dx^b \nonumber \\
&&R_4 \rightarrow R_4~~~~,~~~~~~ 
R_e \rightarrow -R_e~~~~,~~~~~~ 
\sqrt{g} \,d^Dx \rightarrow
\,\sqrt{g} \,d^Dx \label{a11a} 
\end{eqnarray}
where $R_4=g^{\mu\nu}R_{\mu\nu}$ stands for the 
4-dimensional part of the curvature scalar and 
$R_e=g^{ab}R_{ab}$
stands for the extra dimensional part of the curvature scalar and
\begin{eqnarray}
\mu\nu=0,1,2,3~~~~,~~~~~~ 
a,b=4^\prime,5^\prime,....D^\prime-1,4^{\prime\prime}, 
5^{\prime\prime},.....,D^{\prime\prime}-1 \nonumber
\end{eqnarray}
It is evident that the extra 
dimensional part of the gravitational action, that is, 
\begin{equation}
S_{Re} = \frac{1}{16\pi\,G}\int 
\sqrt{g} d^Dx 
\,R_e
\label{a11aa}
\end{equation}
is forbidden by (\ref{a11a}). So only 
\begin{equation}
S_{R4} = \frac{1}{16\pi\,G}\int 
\sqrt{g} \,R_4\,d^Dx \label{a3a}
\end{equation}
may survive. In other words the requirement of the invariance of the 
action under (\ref{a10a}) and 
(\ref{a10b}) separately insures the vanishing of the bulk cosmological 
constant while the requirement of the invariance of the action under 
the simultaneous applications of (\ref{a10a}) and (\ref{a10b}) insures the 
vanishing extra dimensional curvature scalar.

Now I take the discrete symmetry in 
(\ref{a7}) ( or (\ref{a10a}) and (\ref{a10b}) ) be a realization of a 
reflection symmetry in extra 
dimensions and study its 
implications. 
The simplest setup is to realize 
(\ref{a10a}) and (\ref{a10b}) by two reflections in two extra dimensions. 
To be more specific consider 
the following metric ( where 4-dimensional Poincare invariance is taken 
into account \cite{Rubakov}) 
\begin{eqnarray}
ds^2&=&
\Omega_1(y)\Omega_2(z)
g_{\mu\nu}(x)\,dx^\mu dx^\nu\,+ 
\Omega_1(y)g_{AB}(w)\,dx^A dx^B\,+\,
\Omega_2(z)g_{CD}(w)\,dx^C dx^D
\label{a13} \\
\mbox{where}&&~~~
x=x^\mu~,~~y=x^A~,~~z=x^C~,~~w=y,z \nonumber \\
&&\mu,\nu=0,1,2,3~~;~~~~ 
A,B\,=\,4^\prime,5^\prime,....D^\prime-1~~;~~
~C,D\,=\,4^{\prime\prime},5^{\prime\prime},....D^{\prime\prime}-1 
\label{a13aa} \\
&&D^\prime=2(2n+1)~~,~~~D^{\prime\prime}=2(2m+1)~~~~~~n,m=1,2,3,.....
\label{a13ab} 
\end{eqnarray}
and $\Omega_1(y)$, $\Omega_2(z)$
are odd functions of $y$, 
$z$; respectively,  under some 
reflection; and 
$\tilde{g}_{AB}$, 
$\tilde{g}_{CD}$, are even functions of $y$, $z$. For simplicity we 
assume that $\Omega_1$ and 
$\Omega_2$, each depends only on one dimension, that is, 
\begin{equation}
\Omega_1(y)=\Omega_1(y_1)~~~\mbox{and}~~~~ 
\Omega_2(z)=\Omega_2(z_1) \label{a13a}
\end{equation}
where $y_1$ is one of $x^A$ and $y_1$ is one of 
$x^C$. For definiteness one may assume that 
$y_1=x^A=x^{4^\prime}$ and $z_1=x^C=x^{4^{\prime\prime}}$. 
In other words
$y_1=x^A=x^{4^\prime}$ and $z_1=x^C=x^{4^{\prime\prime}}$ are taken as the 
directions 
where $\sqrt{g}\,d^Dx$ changes sign under 
(a set of) spacetime reflections in that direction. 
The volume element and the 
curvature scalar corresponding to (\ref{a13}) are
\begin{eqnarray}
&&\sqrt{g}\,d^Dx\,=\,
\Omega_1^{2n+1}(y_1)
\Omega_2^{2m+1}(z_1)
\sqrt{\tilde{g}}\,d^Dx  \label{a14} \\
R&=&
(\Omega_1\Omega_2)^{-1}
[R_4+\tilde{R}_e
-(D-1)(
\tilde{g}^{4^\prime4^\prime}
\frac{d^2(ln(\Omega_1)}{dy_1^2} 
+\tilde{g}^{4^{\prime\prime}4^{\prime\prime}}
\frac{d^2(ln(\Omega_2)}{dz_1^2}) \nonumber \\
&&-\frac{(D-1)(D-2)}{4}
(\tilde{g}^{4^\prime4^\prime}
(\frac{d\,ln(\Omega_1)}{dy_1})^2
+\tilde{g}^{4^{\prime\prime}4^{\prime\prime}}
(\frac{d\,ln(\Omega_2)}{dz_1})^2)] \label{a15} \\
&& \mbox{where}~~~~D=D^\prime+D^{\prime\prime}-4= 
2(2n+1)+2(2m+1)-4\,=\,4(n+m) \nonumber \\
&&~~~~\tilde{g}^{MN}=\Omega_1(y_1) \Omega_2(z_1)\,g^{MN} 
~~~\tilde{g}^{4^{\prime}4^{\prime}}=\Omega_2(z_1) 
\,g^{4^{\prime}4^{\prime}}
~,~~~~\tilde{g}^{4^{\prime\prime}4^{\prime\prime}}=\Omega_1(y_1) 
\,g^{4^{\prime\prime}4^{\prime\prime}} \label{a15a} 
\end{eqnarray}
$R_4(x)=g^{\mu\nu}R_{\mu\nu}$ and $\tilde{R}_e$ are the curvature scalars 
of the metrics;
$g_{\mu\nu}(x)\,dx^\mu\,dx^\nu$ and 
$\tilde{g}_{AB}(y,z)\,dx^A\,dx^B$+
$\tilde{g}_{CD}(y,z)\,dx^C\,dx^D$
=$\Omega_2^{-1}(z_1)g_{AB}(y,z)\,dx^A\,dx^B$+
$\Omega_1^{-1}(y_1)g_{CD}(y,z)\,dx^C\,dx^D$; respectively.
The action corresponding to (\ref{a14}) and (\ref{a15}) is
\begin{eqnarray}
S_R &=& \frac{1}{16\pi\,G}\int 
\Omega_1^{2n}(y_1)
\Omega_2^{2m}(z_1)
\sqrt{\tilde{\tilde{g}}}
\,d^Dx\,\tilde{R} 
   \label{a17} \\
&&\mbox{where}~~~~~~~~\tilde{R}=[R_4+\tilde{R}_e
-(D-1)(\tilde{g}^{4^\prime4^\prime}\frac{d^2(ln(\Omega_1)}{dy_1^2} 
+\tilde{g}^{4^{\prime\prime}4^{\prime\prime}}
\frac{d^2(ln(\Omega_2)}{dz_1^2}) 
\nonumber \\
&&-\frac{(D-1)(D-2)}{4}
(\tilde{g}^{4^\prime4^\prime}
(\frac{d\,ln(\Omega_1)}{dy_1})^2
+\tilde{g}^{4^{\prime\prime}4^{\prime\prime}}
(\frac{d\,ln(\Omega_2)}{dz_1})^2)] \label{a17a} \\
&&~\mbox{and}~~~~\tilde{\tilde{g}}=
det(g_{\mu\nu})det(g_{AB})det(g_{CD}) \nonumber
\end{eqnarray}
One notices that all terms in $\tilde{R}$ in Eq.(\ref{a17}) except $R_4$ are odd  
either in $y_1$ or in $z_1$ provided that 
$\Omega_1$ is odd ( about some 
point) in $y_1$ and  
$\Omega_2$ is odd ( about some 
point) in $z_1$ and all other terms in (\ref{a17}) are even. 
So all terms in (\ref{a17}) except the  
$R_4$ contribution of $\tilde{R}$ 
vanish after integration. In other words the symmetry imposed ( which 
makes $\Omega_{1(2)}$ odd in $y_{1(z_1)}$ ) guarantees the absence of 
cosmological constant.
For 
example consider 
\begin{equation}
\Omega_1\,=\,\cos{k_1\,x_{5^\prime}}~,~~~~
\Omega_2\,=\,\cos{k_2\,x_{6^{\prime\prime}}} \label{a18}
\end{equation}
Because $\Omega_1$, $\Omega_2$ in (\ref{a18}) are 
odd under the parity operator about 
the point, 
$k_{1(2)}\,x_{5^\prime(6^{\prime\prime})}\,=\,\frac{\pi}{2}$
defined by  
\begin{equation}
k_{1(2)}\,x_{5^\prime(6^{\prime\prime})}
\,
\rightarrow\,\pi-
k_{1(2)}\,x_{5^\prime(6^{\prime\prime})}
\label{a15a}
\end{equation}
and $\frac{d^2(ln(\Omega_{1(2)})}{dy_1(z_1)^2})$ and  
$(\frac{d\,ln(\Omega_{1(2)})}{dy_1(z_1)})^2$ are even hence the 
$\Omega_{1(2)}$ dependent terms in 
(\ref{a17a}) are odd. By the same reason $\tilde{R}_e$ is odd as well. 
So the only even term in 
$\tilde{R}$ is $R_4$. So there is no contribution to the cosmological from 
the bulk cosmological 
constant or from the extra dimensional part of the curvature scalar.  One may 
consider other types of 
spaces as well; for example one may take 
the parity operator be defined 
by $x_{D-1}\rightarrow\,-x_{D-1}$
about 
the point $x_{D-1}=0$ and
either of $\Omega_{1(2)}$ or 
both 
of them change sign under the parity operator (for example, as 
$\Omega=\sin{kx_{D-1})}$).  
In fact one may 
consider a more restricted parity transformation which 
effectively corresponds to the interchange of two branes in the 
$x_{D-1}$-direction. For example one may take 
some dimensions, say the $x_{D-1}$'th dimension, be identified by 
the closed line interval described by $S^1/Z_2$ so that 
$\Omega\,=\,\cos{|k\,x_{D-1}|}$, and 
there are two branes located at 
$x_{D-1}=0$ and $kx_{D-1}=\pi$. Then the transformation in (\ref{a7}) is 
effectively induced by the interchange of the two branes. 

The transformation rule for the Lagrangian
under the requirement of the invariance of the action functional (where 
the metric (\ref{a13}) is considered for simplicity) 
\begin{eqnarray}
S_L &=& \int 
\sqrt{g}
\,d^Dx\,{\cal L}
= \int 
\Omega_1^{2n+1}(y_1)
\Omega_2^{2m+1}(z_1)
\sqrt{\tilde{\tilde{g}}}
\,d^Dx\,{\cal L} \label {a19} \\
&&\mbox{where}~~~~~\tilde{\tilde{g}}=
det(g_{\mu\nu})det(g_{AB})det(g_{CD}) \nonumber
\end{eqnarray} 
under 
Eq.(\ref{a7}) ( or under (\ref{a10a}) and (\ref{a10b}) ) results in
\begin{equation}
{\cal L} \,\rightarrow\,-{\cal L} ~~~~~~\mbox{as}~~~~~~
g_{MN}\rightarrow -\,g_{MN}
~~~~~~~~~~
M,N=0,1,2,3,...... D-1 
\label{a23}
\end{equation}
which is similar to the condition obtained in 
\cite{Erdem}. To be more specific we consider the metrics 
of the form of  
Eq.(\ref{a13}). Then (\ref{a23}) becomes 
\begin{equation}
{\cal L} \,\rightarrow\,-{\cal L} ~~~~~~\mbox{as}~~~~~~
 \Omega_1\,\rightarrow\,-\Omega_1~~~~~~\mbox{and/or}~~~~~~
 \Omega_2\,\rightarrow\,-\Omega_2 \label{a23a}
\end{equation}
After one considers 
the kinetic 
part of the Lagrangian for the scalar fields 
\begin{equation}
2{\cal L}_k= g_{MN}(\partial_M\phi)^\dagger\partial_N\phi=
\Omega_1\Omega_2
g_{\mu\nu}\,
\partial_\mu\phi)^\dagger\partial_\nu\phi+
 \Omega_1
g_{AB}(\partial_A\phi)^\dagger\partial_B\phi+
\Omega_2
g_{CD}(\partial_C\phi)^\dagger\partial_D\phi \label{a24}
\end{equation}
one notices that only the 4-dimensional part of (\ref{a24})
transforms as in the required form, (\ref{a23}) under both of 
$\Omega_{1(2)}\rightarrow
-\Omega_{1(2)}$. So the 
extra dimensional piece of the kinetic Lagrangian for scalar fields is 
forbidden by this symmetry. In other words the extra dimensional part of 
the kinetic Lagrangian vanishes after integration. The scalar field is 
allowed to transform as 
\begin{equation}
\phi\,\rightarrow\,\pm\phi \label{b10}
\end{equation}
If adopt the plus sign in (\ref{b10}) then no potential term is allowed 
in the bulk (if we impose the symmetry) while the terms localized on 
branes may be allowed. 
However introducing potential terms in the bulk 
is not problematic once the symmetry is identified by reflections 
in extra dimensions because these terms 
cancel out 
after integration  over the directions where the volume element is odd 
under these reflections. 
Therefore 
such terms are not dangerous and no restriction 
is put on them in this set-up while some 
restrictions were obtained for such terms in the case of
\cite{Erdem}. Hence the only term which may survive after integration over 
the extra dimensions is the 4-dimensional piece of the kinetic term and it 
does not contribute to the cosmological 
constant since it depends on 4-dimensional coordinates non-trivially. 
So the realization of the symmetry introduced here gives more freedom for 
model building than its previous realization which introduced some 
constrains on the form of the potential terms and the dimensions where 
they may live. 
Similar conclusions are valid 
for the vector fields as well. The case of fermion fields is more 
involved. The potential term 
of the fermionic Lagrangian is not allowed in the bulk ( i.e. 
in each of the $2(2n+1)$ dimensional spaces) by 
Eqs.(\ref{a10a},\ref{a10b}). 
However if 
Eqs.(\ref{a10a},\ref{a10b})are identified as the results of reflections 
in extra dimensions as given 
in (\ref{a15a}) then the potential terms 
cancel out after 
integration 
because they are even 
under (\ref{a15a}). So 
potential terms do not pose a problem. Kinetic term 
of the fermionic Lagrangian 
is neither odd nor even under the 
separate applications of  
Eqs.(\ref{a10a},\ref{a10b}) so it 
does not 
seem to cancel out after integration. To see this better consider 
the specific case where the metric is in the form of (\ref{a13}) 
and $\Omega_{1(2)}$ are given as in (\ref{a18}) with 
\begin{eqnarray}
g_{MN} &=& \Omega\;\eta_{MN} ~~~~\mbox{where}~~~~
\Omega= \begin{array}{ll} 
\Omega_1(u_1)\Omega_2(u_2) 
&~\mbox{for}~~~~M,N=\mu,\nu \\
\Omega_1(u_1)\ &~\mbox{for}~~~~M,N=A,B \\
\Omega_2(u_2)\ &~\mbox{for}~~~~M,N=C,D  \end{array} 
\label{a24a} 
\end{eqnarray}
where $\mu$,$\nu$, $A$, $B$, $C$, $D$ stand for the coordinate indices 
defined in Eq.(\ref{a13aa});
$u_{1(2)}$ stands for 
$k_{1(2)}\,x_{5^\prime(6^{\prime\prime})}$, and 
$\eta_{MN}$ is the D-dimensional flat metric containing the 
usual 4-dimensional 
Minkowski metric. 
The corresponding Lagrangian and action 
functionals are
\begin{eqnarray}
&&{\cal L}_{fk}= 
i\bar{\psi}\Gamma^\mu\partial_\mu\psi+
i\bar{\psi}\Gamma^{a1}\partial_{a1}\psi+
i\bar{\psi}\Gamma^{a2}\partial_{a2}\psi  \label{b14} \\
&&\mbox{where}~~~~~
\Gamma^\mu= 
[(\,\cos{\frac{u_1}{2}}\,\tau_3\,+\,i\,\sin{\frac{u_1}{2}}\,\tau_1\,)
(\,\cos{\frac{u_2}{2}}\,\tau_3\,+\,i\,\sin{\frac{u_2}{2}}\,\tau_1\,)]^{-1}
\otimes\,\gamma^{\tilde{\mu}} \nonumber \\
&&~~
\Gamma^{a1(2)}= 
(\,\cos{\frac{u_{1(2)}}{2}}
\,\tau_3\,+\,i\,\sin{\frac{u_{1(2)}}{2}}\,\tau_1\,)
^{-1}
\otimes\,\gamma^{\tilde{a1(2)}} \nonumber \\
&&\{\Gamma^M,\Gamma^N\}=2\,g^{MN}~~~~~,~~~~
\{\gamma^{\tilde{M}},\gamma^{\tilde{N}}\}=2\,\eta^{\tilde{M}\tilde{N}} 
~~~~~~~M,N=\mu,a1,a2 \label{b14a} \\
S_{Lfk} &=& \int 
\sqrt{g}
\,d^Dx\,{\cal L}_{fk}
= \int 
\Omega_1^{2n}(y_1)
\Omega_2^{2m}(z_1)
\sqrt{\tilde{\tilde{g}}}
\,d^Dx\,{\cal L}^\prime_{fk}
 \label {b15} \\
&&\mbox{where}~~~~{\cal L}^\prime_{fk}= 
{\cal L}^\prime_{fk}= 
i\bar{\psi}\Gamma^{\prime\mu}\partial_\mu\psi+
i\bar{\psi}\Gamma^{\prime a1}\partial_{a1}\psi+
i\bar{\psi}\Gamma^{\prime a2}\partial_{a2}\psi  
\label{b16} \\
&&\Gamma^{\prime\mu}=\Omega_1\Omega_2
\Gamma^{\mu}=
\cos{u_1}\cos{u_2}[(\,\cos{\frac{u_1}{2}}\,\tau_3\,+\,i\,\sin{\frac{u_1}{2}}\,\tau_1\,)
(\,\cos{\frac{u_2}{2}}\,\tau_3\,+\,i\,\sin{\frac{u_2}{2}}\,\tau_1\,)]^{-1}
\otimes\,\gamma^{\tilde{\mu}} \nonumber \\
&&~~\Gamma^{\prime a1(2)}= 
\Omega_{1(2)}\Gamma^{a1(2)}= 
\cos{u_{1(2)}}
(\,\cos{\frac{u_{1(2)}}{2}}\,\tau_3\,+\,i\,sin{\frac{u_{1(2)}}{2}}\,\tau_1\,)
^{-1}
\otimes\,\gamma^{\tilde{a1(2)}} 
\label{b16a} 
\end{eqnarray}
where $\gamma^{\tilde{A}}$ are the usual gamma matrices 
corresponding to $\eta_{MN}$ in (\ref{a24a});
$\tau_1$, $\tau_3$ are the Pauli sigma matrices; 
$\otimes$ denotes 
tensor product. Notice that the number of spinor components for the 
fermions and hence the size of gamma matrices are doubled by the 
introduction of the Pauli sigma matrices in (\ref{b14a}) and (\ref{b16a}). 
This choice is more advantageous than the gamma matrices containing the 
standard vielbeins involving 
$\sqrt{g_{MN}}\propto \sqrt{\Omega}=\sqrt{\cos{u_{1(2)}}}$ 
since 
$\sqrt{\cos{u_{1(2)}}}$ is ill defined under (\ref{a23a}) while the gamma 
matrices introduced above do not pose such a problem.
One 
notices that (\ref{b16}) is multiplied by $\pm\,i$ under \begin{equation}
u_{1(2)}\rightarrow\,\pi-u_{1(2)} \label{b17}
\end{equation}
so the argument of (\ref{b16}) is neither odd nor even under (\ref{b17}) 
on contrary to the scalar 
case, (\ref{a24}). Hence at first sight it seems that the method employed 
here 
to make possible extra dimensional contribution from the fermionic kinetic 
term does not work. However one notices that 
$\cos{u_{1(2)}}(\,\cos{\frac{u_{1(2)}}{2}}\,\tau_3\,
+\,i\,\sin{\frac{u_{1(2)}}{2}}\,
\tau_1\,)^{-1}$ 
is odd under a parity operation about the point $u_{1(2)}=2\pi$ defined 
by 
\begin{equation}
u_{1(2)}\rightarrow\,2\pi-u_{1(2)} \label{b18}
\end{equation}
and the other terms in (\ref{b15}) are even under (\ref{b18}) so that 
$S_{Lfk}$ vanishes after integration. Therefore if a fermion lives in 
the whole bulk then its contribution to the vacuum energy ( and hence to 
the cosmological constant) is zero if adopt the spaces (of the 
form of (\ref{a13}) and (\ref{a18})) whose volume 
elements are odd under spacetime reflections (of some of the extra 
dimensions). If 
one wants to avoid this result 
then the fermions 
must be confined into a subspace where (\ref{b16}) is invariant under 
(\ref{a10a}) and (\ref{a10b}), that is, the fermions must be localized 
in the directions 
(e.g. $y_1$ and/or $z_1$ in (\ref{a14}) ) 
where the volume element is odd under their
reflections
so 
that the fermions live in a $4(n+m)-1$ or $4(n+m)-2$ dimensional subspace 
of the bulk. 

Now we want to point out the relation between this scheme and the 
E-parity model of Linde \cite{Linde}. 
In Linde's model the 
total universe consists of two universes; the usual one 
and ghost particles universe. The corresponding action functional is 
taken as 
\begin{eqnarray}
S= N\,\int\,d^4x\,d^4y\,
\sqrt{g(x)}
\sqrt{g(y)}[ 
\frac{M_{Pl}^2}{16\pi}\,R(x)+{\cal L}(\psi(x))-
\frac{M_{Pl}^2}{16\pi}\,R(y)-{\cal L}(\hat{\psi}(y))]
\label{b25}
\end{eqnarray}
where $\psi$ and $\hat{\psi}$ stand for the usual particles and ghost 
particles, $R(x)$, $R(y)$ are the scalar curvatures of the usual and the 
ghost parts of the universe with the metric tensors $g_{\mu\nu}$ and 
$\hat{g}_{\mu\nu}$; respectively. If one imposes the symmetry 
\begin{eqnarray}
&&P:~ g_{\mu\nu}\, \leftrightarrow\, \hat{g}_{\mu\nu}~,~~~~~P:~ 
\psi\,\leftrightarrow 
\,\hat{\psi} \label{b26} \\
\end{eqnarray}
then any constant which may be induced through Lagrangians is canceled by 
the symmetry so that the vacuum energy hence the cosmological constant is 
zero. In this scenario two universes are assumed to be non-interacting 
(which is a rather strong condition). Other variants and refinements of 
this model are proposed 
in \cite{Kaplan,Moffat}. However the main idea of the scheme is 
preserved in these studies as well. So we do not consider them separately. 
I think the symmetry proposed by Linde has some relation with the symmetry 
studied in this paper. The parity-odd part of the actional 
function in this paper is forbidden or cancels out (depending on if you 
just impose the symmetry or identify it as reflection 
in extra dimension(s)). So as long as the vanishing part of the action 
functional 
is concerned, the action functional in this study transforms as in 
Eq.(\ref{b25}). In the present scheme 
\begin{eqnarray}
&&g_{\mu\nu}\rightarrow -\,g_{\mu\nu} 
~~~~~\mbox{implies}~~~~\begin{array}{l}
\mbox{if}~~~~R,{\cal L}\rightarrow\,
-R,-{\cal L}~~~\mbox{and}~~~S\rightarrow\,-S~~~~\mbox{then}~~~S=0\\
\mbox{if}~~~~R,{\cal L}\rightarrow\,
-R,-{\cal L}~~~\mbox{and}~~~S\rightarrow\,S~~~~\mbox{then}~~~S\neq\,0\\
\mbox{if}~~~~R,{\cal L}\rightarrow\,
R,{\cal L}~~~\mbox{and}~~~S\rightarrow\,-S~~~~\mbox{then}~~~S=0\\
\mbox{if}~~~~R,{\cal L}\rightarrow\,
R,{\cal L}~~~\mbox{and}~~~S\rightarrow\,S~~~~\mbox{then}~~~S\neq\,0 
\end{array}
\label{a28} 
\end{eqnarray}
In other words 
the vanishing cosmological constant 
is related to $S\rightarrow\,-S$
in the present study as well.
As long as the cosmological constant is concerned the conclusion of both 
schemes are similar.
The relation between two schemes can 
be seen better if one 
considers two branes in a space respecting this symmetry. 
Let us consider a space whose metric tensor transforms like 
(\ref{a10a}), \ref{a10b}) and whose volume element is odd under 
reflections  in the direction of the
$x_{4^\prime\,4^{\prime\prime}}$'th dimension(s) and forms a closed line 
interval described by $S^1/Z_2$ with 
the metric of the form of (\ref{a13}) where 
$\Omega_{1(2)}\,=\,\cos{|k_{1(2)}\,
x_{4^\prime(4^{\prime\prime})}|}$. Hence 
there are two branes (for each direction) located at 
$x_{4^\prime(4^{\prime\prime})}=0$ and at
$x_{4^\prime(4^{\prime\prime})}=\pi$.
Then under the transformation given in Eqs.(\ref{a10a}) and(\ref{a10b}) 
two branes are interchanged and for the even terms 
in $R$ and ${\cal L}$ 
(e.g. for cosmological constant), 
$S \rightarrow -S$ so that the contribution of the branes cancel each 
other after integration in a way similar to the Linde's model. Of course 
there are  
essential differences between the two models. The space-time in Linde's 
model 
is 4-dimensional and the volume element of the space was taken to be 
not effected by the symmetry while the space-time in the present model 
is higher dimensional and its volume element is odd under 
the symmetry transformation. So in our model the parts of 
$R$, 
${\cal L}$ which are even under the symmetry cancel out to maintain the 
cosmological constant zero while in Linde's model $R$ and ${\cal L}$ ( or 
at least ${\cal L}$) is odd under the symmetry to make the cosmological 
constant zero. In Linde's model symmetry is ad hoc while in the present 
study the symmetry arises from $g_{AB}\rightarrow\,-g_{AB}$, which can be 
identified by reflection symmetry in extra dimensions. 
 
In this study we have studied the symmetry induced by reversal 
of the sign of the metric tensor. We have identified this 
symmetry by reflections in extra dimensions. In 
this way we may find some higher dimensional spaces which 
satisfy the symmetry and forbid both bulk and 4-dimensional 
cosmological constants. We have also discussed the relation between this 
symmetry and the E-parity symmetry of Linde. Another point worth 
to mention is that throughout this study we 
take the gravity 
propagate in the whole extra dimensions while standard 
model 
particles are localized in a brane (or branes) in the bulk so that the 
contribution of the curvature scalar and the Lagrangian terms in the 
bulk which depend on only extra dimensions cancel out while 
the standard model effects survive.




\bibliographystyle{plain}

\end{document}